\documentclass[acmsmall,screen,natbib=false,
nonacm]{acmart}

\RequirePackage[
datamodel=acmdatamodel,
style=acmauthoryear, sortcites=false, ]{biblatex}
\DeclareCiteCommand{\longcite}
  {\usebibmacro{textcite:prenote}}{\usebibmacro{citeindex}\defcounter{maxnames}{99}\usebibmacro{textcite}}
  {\multicitedelim}
  {\usebibmacro{textcite:postnote}}

\newbibmacro*{textcite:prenote}{\usebibmacro{prenote}}

\DeclareCiteCommand{\longcitep}
  {\usebibmacro{prenote}\bibopenbracket}
  {\usebibmacro{citeindex}\defcounter{maxnames}{99}\usebibmacro{cite}}
  {\multicitedelim}
  {\bibclosebracket\usebibmacro{postnote}}

\usepackage[obeyFinal]{todonotes}

\usepackage{multirow}
\usepackage{varwidth}
\usepackage{mathpartir}
\usepackage{subcaption}

\usepackage{mathtools}
\usepackage[nointegrals]{wasysym}
\usepackage{textcomp}
\usepackage{rotating}
\usepackage{cancel}
\usepackage[center]{formal-grammar}

\usepackage{ebproof}

\usepackage{cleveref}

\crefname{floatgrammar}{gra.}{gra.}
\Crefname{floatgrammar}{Gra.}{Gra.}

\usepackage{crossreftools} \renewcommand*{\phi}{\varphi}
\renewcommand*{\epsilon}{\varepsilon}

\newcommand*{\positive}{\boxplus}
\newcommand*{\notmodal}{\pm}

\newcommand*{\sem}[1]{\left \llbracket #1 \right \rrbracket}

\newcommand{\suppress}[1]{}

\NewDocumentCommand{\semfull}{m m m g}{{\llparenthesis{#1}\rrparenthesis}^{#2}_{\mathsf{#3}}\IfNoValueF{#4}{(#4)}} \NewDocumentCommand{\sembox}{m m g}{\semfull{#1}{}{#2}{#3}}
\NewDocumentCommand{\cpsfull}{m m m g}{\sem{#1}^{#2}_{\mathsf{#3}}\IfNoValueF{#4}{(#4)}}
\NewDocumentCommand{\cpsbox}{m m g}{\cpsfull{#1}{\lambda}{#2}{#3}}
\NewDocumentCommand{\cpsL}{m m g}{\cpsfull{#1}{\mathrm{L}}{#2}{#3}}
\newcommand*{\tp}{{t\mkern-2mu p}}
\newcommand*{\mutilde}{\tilde{\mu}}

\global\long\def\rotxc#1{\begin{sideways}#1\end{sideways}}
\global\long\def\invert#1{\hbox{\rotxc{\rotxc{\ensuremath{#1}}}}}
\newcommand*{\llpar}{\mathbin{\invert{\&}}}
\newcommand*{\with}{\mathbin{\&}}

\newcommand*{\defeq}{\mathrel{\overset{\raisebox{-0.4ex}{\ensuremath{{\scriptscriptstyle \textup{def}}}}}{=}}}

\newcommand*{\mumatch}[4]{\mutilde(\iota_1~#1 . #2 \mid \iota_2~#3 . #4)}
\newcommand*{\mustruct}[4]{\mu(\pi_1~#1 . #2 \mid \pi_2~#3 . #4)}
\newcommand*{\unit}{\text{\usefont{U}{bbold}{m}{n}{1}}}
\newcommand*{\modalsep}{\mathrel{\textnormal{\textbrokenbar}}}
\newcommand*{\letin}[4][]{\operatorname{let}\,#2 = #3\,\operatorname{in}_{#1}\,#4}

\newcommand*{\Lbox}{\mathbf{L}^{\typbox}_{\smash{\mathrm{pol}}}}
 \let\Coloneqq\Coloneqqq

\usepackage{perfectcut}
\makeatletter
\newcommand{\command}[3]{\cut@computeBinary@IncreaseHeight {\mskip\cutangleouterskip \nthleft{\cut@n}{\langle}\mskip\cutangleskip}{#2}{\mskip\cutbarskip {\cut@bars{\cut@n}}^{#1}\mskip\cutbarskip}{#3}{\mskip\cutangleskip \nthright{\cut@n}{\rangle}\mskip\cutangleouterskip}}
\makeatother

\newcommand{\bmid}{\mathrel{\boldsymbol{\mid}}}
\newcommand{\typewithpol}[2]{\underset{#2\vphantom{\hat{\mid}}}{#1}}

\newcommand*{\Downshift}{\mathord{\Downarrow}}
\newcommand*{\Upshift}{\mathord{\Uparrow}}

\newcommand*{\valbox}{\mathord{\square}}
\newcommand*{\polbox}{\mathord{\square}}
\newcommand*{\typbox}{\mathord{\Box}}

\newcommand*{\restrict}[2]{\mathopen{} #1 \mathord\mid_{#2} \mathclose{}}

\newcommand*{\semval}[1]{\mathbb{#1}}
\newcommand*{\semstack}[1]{\mathbb{#1}}

\newcommand*{\envheap}[1]{\mathcal{#1}}
\newcommand*{\enveq}[1]{\mathcal{#1}}
\newcommand*{\envstack}[1]{\mathcal{#1}}
\newcommand*{\envmem}[1]{\mathcal{#1}}
\newcommand*{\mem}[2]{(#1; #2)}

\newcommand*{\evalval}[3]{#1 \vDash #2 \Downarrow_v #3}
\newcommand*{\evalstack}[3]{#1 \vDash #2 \Downarrow_s #3}
\NewDocumentCommand{\redcmd}{s m m m m}{\left(#2 \vDash #3\right) \IfBooleanTF{#1}{\rightsquigarrow^*}{\rightsquigarrow} \left(#5 \vDash #4\right)}

\newcommand*{\red}{\mathrel{\vartriangleright}}

\newcommand*{\semV}{\semval{V}}
\newcommand*{\semW}{\semval{W}}
\newcommand*{\semS}{\semstack{S}}
\newcommand*{\envH}{\envheap{H}}
\newcommand*{\envE}{\enveq{E}}
\newcommand*{\envF}{\enveq{F}}
\newcommand*{\envS}{\envstack{S}}
\newcommand*{\envM}{\envmem{M}}

\newcommand*{\subst}[2]{[{#1}/{#2}]}
\newcommand*{\addsubst}[2]{,{#1}/{#2}}

\newcommand*{\rulename}[1]{$\mkern-6mu\scriptstyle (#1)$}
\newcommand*{\rulenameleft}[1]{\rulename{#1{\vdash}}}
\newcommand*{\rulenameright}[1]{\rulename{{\vdash}#1}}

\definecolor{dark-red}{rgb}{0.6,0,0}
\definecolor{dark-blue}{rgb}{0,0.0,0.6}
\definecolor{blue}{rgb}{0,0,0.8}
\hypersetup{
unicode=true,pdfusetitle,
 bookmarks=true,bookmarksnumbered=false,bookmarksopen=false,
 breaklinks=true,pdfborder={0 0
 0},pdfborderstyle={},backref=page,colorlinks=true,
 ocgcolorlinks,linkcolor={dark-red},citecolor={dark-blue},urlcolor={blue}}

\title
[S4 modal sequent calculus as intermediate logic and intermediate language]
{S4 modal sequent calculus as intermediate logic\texorpdfstring{\\}{ }and intermediate language}

\thanks{\today. Paper accepted for PEPM 2026 in the ``short paper''
  category.}

\author{Jean Caspar}
\orcid{0009-0003-2732-7628}
\email{jean.caspar@ens.psl.eu}
\affiliation{ \institution{École Normale Supérieure --- PSL},
  \institution{INRIA} \city{Paris} \country{France} }

\author{Guillaume Munch-Maccagnoni}
\orcid{0009-0001-7914-7165}
\email{guillaume.munch-maccagnoni@inria.fr}
\affiliation{
  \institution{INRIA, LS2N CNRS}
  \city{Nantes}
  \country{France}
}

\begin{document}

\begin{abstract}
  In this short paper, we advocate for the idea that
  continuation-based intermediate languages correspond to intermediate
  logics. The goal of intermediate languages is to serve as a basis
  for compiler intermediate representations, allowing to represent
  expressive program transformations for optimisation and compilation,
  while preserving the properties that make programs compilable
  efficiently in the first place, such as the ``stackability'' of
  continuations. Intermediate logics are logics between intuitionistic
  and classical logic in terms of provability.

  Second-class continuations used in CPS-based intermediate languages
  correspond to a classical modal logic S4 with the added restriction
  that implications may only return modal types. This indeed
  corresponds to an intermediate logic, owing to the
  Gödel-McKinsey-Tarski theorem which states the intuitionistic nature
  of the modal fragment of S4.

  We introduce $\Lbox$, a three-kinded polarised sequent calculus for
  S4, together with an operational machine model that separates a heap
  from a stack. With this model we study a stackability property for
  the modal fragment of S4.

\end{abstract}

\maketitle

\section{Introduction}

\paragraph{Intermediate languages as intermediate logics}
The proof-theoretical study of continuations through the question of
constructivity in classical logic, via the formula-as-type
correspondence between control operators and classical axioms, has
been rich in lessons. It provided, for instance, a better
understanding of control operators and better calculi to model
computing with continuations, including the Curry-Howard
correspondence between sequent calculus and abstract machines.

Historically, compilation has been one of the roots of the concept of
continuation~\citep{Steele:MS}. Various works have applied the lessons
of the proof-theoretical study to the construction of
compilers~\citep{downen_sequent_2016, schuster_compiling_2025}. At the
same time, there is a more general debate on the relevance of
continuations for modern
compilers~\citep{Maurer2017,cong_compiling_2019}.

The notion of continuation used in the intermediate representations of
compilers and the one studied by logicians turned out to be different.
While the (polarised, \emph{i.e.} evaluation-order aware, versions of)
sequent calculi LJ (intuionistic: linear in the conclusions) and LK
(classical: non-linear in the conclusions) correspond respectively to
linearly-used and non-linear continuations, the restrictions on
continuations that appear in intermediate representations seem to
correspond---by Curry-Howard---to an intermediate logic between
intuitionistic and classical sequent calculus.
This point is made the most clearly by \longcite{downen_sequent_2016}:
\begin{quote}
\hspace*{1em}\emph{\emph{[T]}\kern-0.05em he logic that comes out of Sequent Core lies in between
\emph{[LK and LJ]}: sometimes there can only be one conclusion, and
sometimes there can be many. Furthermore \emph{[...]} Sequent Core
still captures a similar notion of purity from the $\lambda$-calculus.
This demonstrates that there is room for more subtle variations on
intuitionistic logic that lie between the freedom of LK and the purity
of LJ.}
\end{quote}
However, this intermediate character of continuations used for
compilation remained at odds with the proof-theoretic treatment until
\longcite{cong_compiling_2019}, starting from an experiment to
reproduce results from \citet{Maurer2017}, proposed to capture the
``second-class'' character of continuations with a simple type system,
based on ealier work on second-class values
\longcitep{osvald_gentrication_2016}. The language and type system
from \citet{cong_compiling_2019} proved amenable indeed to be modelled
via a sequent calculus for the modal logic S4 in the first author's
MSc thesis~\citep{Caspar2025MScThesis}, providing a logical
characterisation of second-class values.

The calculus of \citet{cong_compiling_2019} distinguishes, thus,
second-class (restricted) value from first-class (unrestricted) ones.
The restrictions on second-class values ensure that they cannot escape
from their defining scope: they cannot be returned from functions, and
cannot be referred to from first-class values. In particular, a
closure referring to a second-class value is itself second-class. This
ensures \citep{osvald_gentrication_2016} that second-class values can
be allocated on the stack, and that the stack can be freed when a
function accepts a first-class argument.
Continuations introduced by the compiler following
\citet{cong_compiling_2019} are second class, ensuring they they can
be allocated in a stack-like fashion. Nevertheless, from a
Curry-Howard perspective, their CPS translation presents classical
features: continuations can be duplicated or erased to a certain
extent, relaxing the strict linear use of continuations in CPS models
of intuitistic logic.

More recently, \longcite{schuster_compiling_2025} challenged our
preconceptions by using as a basis for efficient compilation a
one-sided polarised classical sequent calculus---originally intended
in \citet{munch-maccagnoni_focalisation_2009} as a calculus for
Girard's perfectly symmetric classical
logic~\longcitep{girard_new_1991,danos_new_1997}, and expressing
unrestricted \emph{call/cc} operators. Since it does not feature a
priori restrictions on the use of continuations, the trick is to rely
on a dynamic use of linearity in the back-end, using
reference-counting with re-use of linearly-used memory cells. As they
note, \emph{``since continuation frames are typically used linearly, we
hence get fast code for stack-like usage, even though we do not
maintain a stack.''}

We believe that there still are advantages in making the intermediate
representations aware of the effective linearity of continuations, in
order to better preserve (e.g. during code transformations for
optimisation) and make use of (e.g. using hardware stack support)
these properties which yield an efficient compilation strategy.

\paragraph{A classical S4 sequent calculus as an intermediate representation}

Building upon the work by \longcite{cong_compiling_2019}, this paper
reports on some results from the first author's MSc
thesis~\citep{Caspar2025MScThesis} which provided an investigation
into second-class continuations with a polarised version of
(classical) S4 modal sequent calculus, the calculus $\Lbox$. One of
its main results that motivates this short paper, though we do not
detail it here, is its analysis of \citet{cong_compiling_2019} though
the lens of modal logic:
\begin{center}
\fbox{
\begin{minipage}{0.7\linewidth}{
\emph{The CPS translation in
  \citet{cong_compiling_2019} factors into an interpretation of
  their calculus into $\Lbox$, followed by a CPS translation of
  $\Lbox$.}}
\end{minipage}
}
\end{center}\vspace*{0.5ex}
This extends a similar decomposition of CPS translations for the
$\lambda\mu$-calculi via the one-sided classical sequent
calculus~\cite{danos_new_1997,Ogata2000,Lau02PhD,munch-maccagnoni_syntax_2013}.
In this short paper, we omit the CPS translation for $\Lbox$ and establish
properties of its operational semantics directly.

The intuitionistic S4 modal logic has found a wide range of
applications in programming languages, including staged compilation
\longcitep{davies_modal_2001}, contextual type theory \longcitep{Nanevski2008},
and information flow \longcitep{miyamoto_modal_2004}.

The \emph{(classical)} S4 modal logic can be seen as mixing classical
logic and intuitionistic logic, owing to the Gödel-McKinsey-Tarski
theorem~\longcitep{godel_interpretation_2001,mckinsey_theorems_1948},
according to which its modal fragment is complete for intuitionistic
provability. Furthermore, if we consider a fragment of classical S4
modal logic obtained by restriction to functions whose return types
are modal, we obtain an intermediate logic between classical and
intuitionistic. We are indeed affecting the classical character of S4
with this restriction---the modal fragment is not altered---but
limited forms of classical axioms are still provable. The intuition is
that regular types are second class (cannot escape their defining
scope, in programming terms), whereas modal types are first class.

\paragraph{A theory of second-class continuations via S4}
In this paper, we present the polarised sequent calculus $\Lbox$ for
the modal logic S4, with three polarities ($\polbox,+,-$).
We also present a machine-like semantics with features similar to
those of \citet{cong_compiling_2019}: we divide the memory into two
parts, an ordinary heap for modal values, which can only refer to
modal variables, and another part which is stack-like and gets freed
every time we evaluate a modal covariable. The stack is not freed when
returning a value with a non-modal type.

Then, if we restrict S4 to functions with a modal return type, thereby
fully implementing the second-class restriction, we recover the
stackability property of \citet{cong_compiling_2019}. The result
suggests that metatheoretical results for second-class continuations
follow in a straightforward manner from those for S4.

\section{System \texorpdfstring{$\Lbox$}{L□pol}}

In this section, we introduce a polarised sequent calculus for
classical S4 logic with a term syntax called $\Lbox$, which extends
the polarised classical
L-calculus~\citet{munch-maccagnoni_focalisation_2009}.

\paragraph{Brief recapitulation of the polarised L-calculus.}
The grammar of the calculus is formed by commands $c$ of the form
$\command{}{t}{e}$, which are formed with an expression $t$ in an
environment $e$ (a reification of an evaluation context). Environments
can bind variables; for example, as usual, the program
``$\letin{x}{t}{c}$'' can be represented as $\command{}{t}{\mutilde x
  . c}$: it evaluates $t$ and then executes $c$ with the value of $t$
being bound to the variable $x$.

Other constructs are available in $\Lbox$, for instance pairs $(t, u)$
with pattern-matching on such pairs using the environment $\mutilde(x,
y) . c$. Thus, the program $\letin{(x, y)}{t}{c}$ is represented with
$\command{}{t}{\mutilde(x, y) . c}$, which evaluates $t$ to a value
and then executes $c$ with the value of $t$ bound to $(x,y)$, if this
value indeed decomposes into a pair.
We also have a unit value $()$ on which we can match with $\mutilde
().c$ and sum types with injections $\iota_i\,t$ and the match
construct $\mumatch{x}{c}{y}{c'}$.

The type system corresponds to classical logic and follows the
formulae-as-types interpretation for call/cc-style control
operators~\citep{griffin_formulae-as-type_1989}. The current
continuation (or, here, the environment) is captured with the
construction $\mu\alpha . c$: this expression captures the current
environment, binds it to a ``covariable'' $\alpha$, and continues with
$c$.

Following the linear analysis of classical
logic~\citep{girard_new_1991,danos_new_1997}, the apparent
non-joinable critical pair, if $\alpha$ and $x$ are fresh,
\[c \leftarrow \command{}{\mu\alpha . c}{\mutilde x . c'} \to c'\]
is resolved by giving priority to the
left-hand-side reduction for ``positive'' types (corresponding to CBV evaluation), and the right-hand-side reduction for
``negative'' types (corresponding to CBN evaluation). Binders
and commands are annotated by a polarity $\epsilon$.

We must also now distinguish which terms may be substituted or not.
Thus, we have two other syntactic classes: values $V$ and co-values
$S$, which corresponds to ``pure'' expressions and environments that
can be substituted. $()$ is a value, and we restrict pairs and
injections such that only $(V, V')$ and $\iota_i\,V$ are allowed as
values. $\mu\alpha^{\epsilon} . c$ is a value when $\epsilon$ is
negative, but not when $\epsilon$ is positive; $\mutilde x^{\epsilon}
. c$ is a co-value if $\epsilon$ is negative, but not when $\epsilon$
is negative. Except for those, every expression introduced before is a
value, and every environment a co-value: variables are values, and
covariables and $\mutilde(x^{\epsilon_1}, y^{\epsilon_2}) . c$ are
co-values. We have the reduction rules:
\begin{align*}
\command{\epsilon}{\mu\alpha^{\epsilon} . c}{S} &\to c\subst{S}{\alpha}\\
\command{\epsilon}{V}{\mutilde x^{\epsilon} . c} &\to c\subst{V}{x}
\end{align*}
The reduction of $\command{\epsilon}{\mu\alpha^{\epsilon} .
  c}{\mutilde x^{\epsilon} . c'}$ is then determined uniquely by the
polarity $\epsilon$. Since the polarity is determined by the type, the
polarities of both sides match if the command is well-typed.

This system supports both CBV and CBN, so in addition to strict pairs
$(V, W)$ we also have lazy records
$\mustruct{\alpha^{\epsilon}}{c}{\beta^{\epsilon'}}{c'}$. This
represents a record with two fields whose computation is not finished:
to access a field, we can use the co-value $\pi_i\,S$, with $S$
representing the environment in which we want to use the value
computed from the field $i$. The reduction
$\command{-}{\mustruct{\alpha_1^{\epsilon_1}}{c_1}{\alpha_2^{\epsilon_2}}{c_2}}{\pi_i\,S}
\to c_i\subst{S}{\alpha_i}$ assigns this co-value to $\alpha_i$ and
computes the field.
We also have co-values representing pairs of co-values $(S, S')$, the
corresponding values are $\mu(\alpha^{\epsilon_1}, \beta^{\epsilon_2})
. c$, i.e. an expression which inspects the current co-value and
retrieves $S$ and $S'$, and co-values consisting of a single value
$[V]$, with opposing values $\mu[x^{\epsilon}] . c$, which implement
negation.

\paragraph{Adding a polarity $\valbox$.}
The system presented so far corresponds to (a fragment of) classical
polarised L-calculus. We do not merely add the $\typbox$ modality of
S4: we consider a new polarity, $\valbox$, which is similar to $+$ in
the sense that it follows a CBV evaluation. Corresponding values are
introduced by $\valbox V$, and co-values by $\mutilde \valbox
x^{\epsilon} . c$. Moreover, modal variables, annotated by $\valbox$,
are special because a modal value can only refer to modal variables;
this is enforced by typing.

What distinguishes a polarity from a modality is that a polarised type
system reflects properties of stability under type
constructions---typically, a positive pair of two modal values is
itself modal. This is reflected in the type system with polarity
tables which describe how the polarity of a type is deduced from its
constituents.\footnote{This idea goes back to
\citet{girard_new_1991,girard_unity_1993}, which indirectly influenced
modal calculi via the Linear-Non-Linear term calculus and
models~\citep{BentonLinearNonLinear} (at least).} In categorical
terms, this amounts to shifting attention from the comonad $\typbox$
to its decomposition into an adjunction, typically with the category
of $\typbox$-coalgebras, and interpeting a deductive system across
several categories related by adjunctions. We will not get into
categorical details here, but our types $A$ with polarity $\polbox$
all enjoy $\typbox A\rightarrow A$.

In the end, we have three polarities: the usual positive and negative
polarities $+,-$, and one additional modal positive polarity
$\polbox$. Positive polarities $+,\polbox$ will be denoted $\positive$
and non-modal polarities $+,-$ will be denoted $\notmodal$.
As we will see, the polarity $\polbox$ behaves almost like $+$, and
shares a lot of constructs with it. However, their behaviour will
differ in the second semantics we give in
\cref{fig:semvalstack,fig:semcmd}.

\begin{grammar}[Syntax of system $\Lbox$][h][gra:syntax]
  \firstcasesubtil{$\epsilon$}{+ \gralt - \gralt \polbox}{Polarities}
  \firstcasesubtil{$\positive$}{+ \gralt \polbox}{Positive polarities}
  \firstcasesubtil{$\notmodal$}{+ \gralt -}{Non-modal polarities}
  \firstcasesubtil{$V, W$}{x \mkern-5mu \gralt \mkern-5mu (V, W) \mkern-5mu \gralt \mkern-5mu \valbox V \mkern-5mu \gralt \mkern-5mu () \mkern-5mu \gralt \mkern-5mu \iota_1~V \mkern-5mu \gralt \mkern-5mu \iota_2~V \mkern-5mu \gralt \mkern-5mu \mu[x^{\epsilon}] . c}{Values}
\otherform{\mustruct{\alpha^{\epsilon_1}}{c_1}{\beta^{\epsilon_2}}{c_2} \mkern-5mu \gralt \mkern-5mu \mu(\alpha^{\epsilon_1}, \beta^{\epsilon_2}) . c \mkern-5mu \gralt \mkern-5mu \mu \alpha^- . c}{}
  \firstcasesubtil{$S, S'$}{\alpha \mkern-5mu \gralt \mkern-5mu \pi_1~S \mkern-5mu \gralt \mkern-5mu \pi_2~S \mkern-5mu \gralt \mkern-5mu \mutilde\valbox x^{\epsilon} . c \mkern-5mu \gralt \mkern-5mu [V]}{Co-values}
    \otherform{(S, S') \mkern-5mu \gralt \mkern-5mu \mutilde () . c \mkern-5mu \gralt \mkern-5mu \mutilde x^{\positive} . c \mkern-5mu \gralt \mkern-5mu \mutilde(x^{\epsilon_1}, y^{\epsilon_2}) . c}{}
    \otherform{\mumatch{x^{\epsilon_1}}{c_1}{y^{\epsilon_2}}{c_2}}{}
  \firstcasesubtil{$t$}{V \mkern-5mu \gralt \mkern-5mu \mu\alpha^{\positive} . c}{Expressions}
  \firstcasesubtil{$e$}{S \mkern-5mu \gralt \mkern-5mu \mutilde x^- . c}{Environments}
  \firstcasesubtil{$c$}{\command{\positive}{t}{S} \mkern-5mu \gralt \mkern-5mu \command{-}{V}{e}}{Commands}
\end{grammar}

We define an operational semantics $\red$ for this system
(\cref{fig:opsem}). It only reduces at the toplevel, and it is
deterministic: if $c \red c_1$ and $c \red c_2$, then $c_1 =
c_2$.

\begin{figure}[h]
$\begin{array}{ccc}
    \command{\epsilon}{\mu\alpha^\epsilon . c}{S} &\red& c\subst{S}{\alpha} \\
    \command{\epsilon}{V}{\mutilde x^\epsilon . c} &\red& c\subst{V}{x} \\
    \command{\polbox}{()}{\mutilde () . c} &\red& c \\
    \command{\epsilon}{(V, W)}{\mutilde(x^{\epsilon_1}, y^{\epsilon_2}) . c} &\red& c\subst{V}{x\addsubst{W}{y}} \\
    \command{\epsilon}{\iota_i~V}{\mumatch{x_1^{\epsilon_1}}{c_1}{x_2^{\epsilon_2}}{c_2}} &\red& c_i\subst{V}{x_i} \\
    \command{\polbox}{\valbox V}{\mutilde\valbox x^{\epsilon} . c} &\red& c\subst{V}{x} \\
    \command{-}{\mu[x^{\epsilon}] . c}{[V]} &\red& c\subst{V}{x} \\
    \command{-}{\mu(\alpha^{\epsilon_1}, \beta^{\epsilon_2}) . c}{(S, S')} &\red& c\subst{S}{\alpha\addsubst{S'}{\beta}} \\
    \command{-}{\mustruct{\alpha_1^{\epsilon_1}}{c_1}{\alpha_2^{\epsilon_2}}{c_2}}{\pi_i~S} &\red& c_i\subst{S}{\alpha_i} \\
\end{array}$
\caption{Operational semantics of system $\Lbox$}
\label{fig:opsem}
\end{figure}

\section{Typing}

Types and their associated polarities are defined as follows:
\begin{align*}
  \typewithpol{A}{\varpi(A),}, \typewithpol{B}{\varpi(B)} &
  \typewithpol{\Coloneqq}{=}
  \typewithpol{\unit}{\polbox} \bmid \typewithpol{A \otimes B}{x} \bmid \typewithpol{A \oplus B}{x} \bmid
  \typewithpol{\typbox A}{\polbox} \bmid \typewithpol{\neg A}{-} \bmid \typewithpol{A \with B}{-} \bmid \typewithpol{A \llpar B}{-}
\end{align*}
where $x=\varpi(A) \odot \varpi(B)$, defined further below.
In words, to the connectives of classical linear logic without the
exponentials, we add a modality $\typbox$, representing a strong
comonad. However, the system is not linear (hence the comonad is
written ``$\typbox$'' rather than ``$!$''). Thus, even though $A
\oplus B$ and $A \llpar B$ are logically equivalent and represent a
disjunction, they are not isomorphic and their reduction rules differ.
Similarly, $\otimes$ and $\with$ both represent conjunction, and
$\unit$ is the unit of $\otimes$. Finally, $\neg$ is the (negative)
negation.

Types have a polarity $\varpi(A) = \epsilon$: we have $\varpi(A) = -$ for
negative types, $\varpi(A) = +$ for positive types which are not
modal, and $\varpi(A) = \polbox$ for modal types; essentially coalgebras for $\typbox$. Positive and modal
types correspond to CBV types, and negative ones to
CBN types. The operation $\odot$ on polarities is defined as follows:
\begin{align*}
\epsilon \odot \epsilon' &= \begin{cases}
\polbox &\text{ whenever }(\epsilon, \epsilon') = (\polbox, \polbox)\\
+ &\text{ otherwise.}
\end{cases}
\end{align*}
We note $A_{\epsilon}$ to assert that the type $A$ is such that
$\varpi(A) = \epsilon$.

One can see that there is no function type in this system. The type $A
\to B$, representing the arrow type of call-by-push-value, can be macro-defined as $\neg A \llpar B$, with $\llpar$
representing the negative disjunction; values of type $A \to B$ are
introduced by $\mu(a \cdot \beta) . c \defeq \mu(\alpha, \beta) .
\command{-}{\mu[a] . c}{\alpha}$ --- in CBN, $\lambda x . t$
is defined as $\mu(x \cdot \beta) . \command{-}{t}{\beta}$: the idea
is that a value of type $A \to B$ captures a value of type $A$ and a
continuation of type $B$: the value the function returns must be sent to this
continuation. Co-values of type $A \to B$ are introduced by $V \cdot S \defeq
([V], S)$. The idea is that a co-value of type $A \to B$ consists of a
value of type $V$ on top of a co-value of type $B$, which represent the
co-value on which the result of the function will run: for example, $t~u$
is translated as $\mu \beta^- . \command{-}{t}{u \cdot \beta}$ ($u$ is
a value because in CBN, every expression is translated as a value
of negative type).

Similarly, one can define polarity shifts, which cast the type so that
it becomes positive/negative: the type $\Upshift A \defeq A \otimes
\unit$ is positive: its values are $\Upshift V \defeq (V, ())$ and its
co-values are $\mutilde \Upshift x^{\epsilon} . c \defeq \mutilde
(x^{\epsilon}, u^{\polbox}) . c$, and the type $\Downshift A \defeq
\neg \unit \llpar A$ is negative, with values $\mu \Downshift
\alpha^{\epsilon} . c \defeq \mu(\beta^-, \alpha) . c$ and co-values
$\Downshift S \defeq ([()], S)$.

The modal logic S4 a logic which a modality $\typbox$ satisfying $\typbox(A \to B) \to
\typbox A \to \typbox B$, $\typbox A \to A$ and $\typbox A \to \typbox
\typbox A$.
It also comes with the necessitation rule, saying that if $A$ is
provable without hypothesis then $\typbox A$ is provable without
hypothesis. Here, $\typbox$ follows almost the same rules. However,
mixing modalities and evaluation order (effects) reveals difficulties
associated with a value restriction similarly to
\longcite{curien_theory_2016}: as they explain, having both a co-monad
on a category of values, together with an effect adjunction, would
require the rule ${\vdash}\typbox$ to be restricted to values. Thus,
in $\Lbox$, from a proof of $A \to B$ we cannot in general deduce a
proof of $\typbox A \to \typbox B$, but only $\typbox A \to \typbox
\Upshift B$, due to this value restriction. This peculiarity has been
addressed differently in \citet{curien_theory_2016} by taking the
modality to be essentially $\typbox\Upshift$. Our treatment of the
modality differs from \citet{curien_theory_2016} in two aspects: the
construct $\valbox V$ does not introduce an implicit suspension, and
modal contexts are handled as in \citet{kavvos_dual-context_2017}.
Intuitively that $\typbox$ is product-preserving, as in the semantics
of \citet{kavvos_dual-context_2017} and unlike the modality $!$ of
\citet{curien_theory_2016}, for which our approach would not work.

Contexts $(\Gamma \modalsep \Theta \vdash \Delta)$ are made of three
parts, represented by ordered lists of (co)variables: $\Gamma$,
$\Theta$ and $\Delta$. None of them are linear: they all supports
permutation, contraction and weakening. $\Gamma$ is the usual
``intuitionistic'' part of the context, and contains variables with
their type. $\Theta$ is the modal part, also consisting of variables
and their types: $x : A \in \Theta$ should be seen as $x : \typbox A
\in \Gamma$, but it allows to extract the values behind the modality.
Lastly, $\Delta$ is the classical part, consisting of co-variables
$\alpha : A$, representing continuations. An empty list is represented
as $\diamond$.

Indeed, following the Curry-Howard correspondence for classical logic,
we know that formulas on the right-hand side of the $\vdash$
correspond to continuations variables.
The syntax is essentially a fragment of the two-sided L-calculus
\citet{munch-maccagnoni_focalisation_2009,munch-maccagnoni_note_2017},
with a dual context on the left inspired by the syntax of
\citet{kavvos_dual-context_2017} specific to product-preserving
modalities. Lastly, sequents may also include a distinguished zone
(except for commands judgments), which can be left or right, and
contains at most one formula. Intuitively, the formula in this zone is
the one we are working on.

We consider the typing rules given in \cref{fig:lbox}. There are five
typing judgments:
\begin{itemize}
  \item ``$\Gamma \modalsep \Theta \vdash V : A; \Delta$'' : $V$ is a value of type $A$ in the context $(\Gamma \modalsep \Theta \vdash \Delta)$;
  \item ``$\Gamma \modalsep \Theta \vdash t : A \mid \Delta$'' : $t$ is an expression of type $A$ in the context $(\Gamma \modalsep \Theta \vdash \Delta)$;
  \item ``$\Gamma \modalsep \Theta ; S : A \vdash \Delta$'' : $S$ is a co-value of type $A$ in the context $(\Gamma \modalsep \Theta \vdash \Delta)$;
  \item ``$\Gamma \modalsep \Theta \mid e : A \vdash \Delta$'' : $e$ is an environment of type $A$ in the context $(\Gamma \modalsep \Theta \vdash \Delta)$;
  \item ``$c : (\Gamma \modalsep \Theta \vdash \Delta)$'' : $c$ is a command in the context $(\Gamma \modalsep \Theta \vdash \Delta)$.
\end{itemize}

When writing $\Gamma, \Gamma'$, $\Theta, \Theta'$ or $\Delta,
\Delta'$, we assume that $\Gamma$ and $\Gamma'$, $\Theta$ and
$\Theta'$ and $\Delta$ and $\Delta'$ are disjoint. $\Gamma_{\polbox}$ denotes a context in which all types $A$ are such that $\varpi(A) =
\polbox$. We define renamings $\theta \in \mathfrak{R}(\Gamma
\modalsep \Theta \vdash \Delta \Rightarrow \Gamma' \modalsep \Theta'
\vdash \Delta')$: $\theta$ is a function between variables and
covariables such that if $x : A \in \Gamma$ then $\theta(x) : A \in
\Gamma'$, if $x : A \in \Theta$ then $\theta(x) : A \in \Theta'$ and
if $\alpha : A \in \Delta$ then $\theta(\alpha) : A \in \Delta'$.

\begin{figure*}
\renewcommand{\arraystretch}{2.9}
\arraycolsep=0.9ex
\begin{small}
$\mkern-40mu\begin{array}{cc}
\begin{prooftree}
    \infer0[\rulenameright{\text{Ax}}]{x : A \modalsep \diamond \vdash x : A; \diamond}
\end{prooftree} &
\begin{prooftree}
    \infer0[\rulenameright{\typbox\text{Ax}}]{\diamond \modalsep x : A \vdash x : A; \diamond}
\end{prooftree} \vspace*{-1.5ex}\\
\multicolumn{2}{c}{\begin{prooftree}
    \infer0[\rulenameleft{\text{Ax}}]{\diamond \modalsep \diamond \mid \alpha &: A \vdash \alpha : A}
\end{prooftree}} \\
\multicolumn{2}{c}{\begin{prooftree}
    \hypo{\Gamma \modalsep \Theta \vdash V : A_{\positive} ; \Delta}
    \hypo{\Gamma' \modalsep \Theta' \mid e : A_{\positive} \vdash \Delta'}
    \infer2[\rulename{\text{cut}{\positive}}]{\command{\positive}{V}{e} : (\Gamma, \Gamma' \modalsep \Theta, \Theta' \vdash \Delta, \Delta')}
\end{prooftree}} \\
\multicolumn{2}{c}{\begin{prooftree}
    \hypo{\Gamma \modalsep \Theta \vdash t : A_- \mid \Delta}
    \hypo{\Gamma' \modalsep \Theta' ; S : A_- \vdash \Delta'}
    \infer2[\rulename{\text{cut}{-}}]{\command{-}{t}{S} : (\Gamma, \Gamma' \modalsep \Theta, \Theta' \vdash \Delta, \Delta')}
\end{prooftree}} \\
\begin{prooftree}
    \hypo{c : (\Gamma \modalsep \Theta \vdash \alpha : A_{\positive}, \Delta)}
    \infer1[\rulenameright{\mu\positive}]{\Gamma \modalsep \Theta \vdash \mu\alpha^{\positive} . c : A_{\positive} \mid \Delta}
\end{prooftree} &
\begin{prooftree}
    \hypo{c : (\Gamma \modalsep \Theta \vdash \alpha : A_-, \Delta)}
    \infer1[\rulenameright{\mu-}]{\Gamma \modalsep \Theta \vdash \mu\alpha^- . c : A_- ; \Delta}
\end{prooftree} \\
\begin{prooftree}
    \hypo{c : (\Gamma, x : A_{\positive} \modalsep \Theta \vdash \Delta)}
    \infer1[\rulenameleft{\mutilde\positive}]{\Gamma \modalsep \Theta ; \mutilde x^{\positive} . c : A_{\positive} \vdash \Delta}
\end{prooftree} &
\begin{prooftree}
    \hypo{c : (\Gamma, x : A_- \modalsep \Theta \vdash \Delta)}
    \infer1[\rulenameleft{\mutilde-}]{\Gamma \modalsep \Theta ; \mutilde x^- . c : A_- \vdash \Delta}
\end{prooftree} \\
\begin{prooftree}
    \infer0[\rulenameright{\unit}]{\diamond \modalsep \diamond \vdash () : \unit; \diamond}
\end{prooftree} &
\begin{prooftree}
    \hypo{c : (\Gamma \modalsep \Theta \vdash \Delta)}
    \infer1[\rulenameleft{\unit}]{\Gamma \modalsep \Theta ; \mutilde () . c : \unit \vdash \Delta}
\end{prooftree} \\
\multicolumn{2}{c}{\begin{prooftree}
    \hypo{\Gamma \modalsep \Theta \vdash V : A; \Delta}
    \hypo{\Gamma' \modalsep \Theta' \vdash W : B; \Delta'}
    \infer2[\rulenameright{\otimes}]{\Gamma, \Gamma' \modalsep \Theta, \Theta' \vdash (V, W) : A \otimes B; \Delta, \Delta'}
\end{prooftree}} \\
\multicolumn{2}{c}{\begin{prooftree}
    \hypo{c : (\Gamma, x : A_{\epsilon_1}, y : B_{\epsilon_2} \modalsep \Theta \vdash \Delta)}
    \infer1[\rulenameleft{\otimes}]{\Gamma \modalsep \Theta ; \mutilde(x^{\epsilon_1}, y^{\epsilon_2}) . c : A_{\epsilon_1} \otimes B_{\epsilon_2} \vdash \Delta}
\end{prooftree}} \\
\multicolumn{2}{c}{\begin{prooftree}
    \hypo{\Gamma \modalsep \Theta \vdash V : A_i; \Delta}
    \infer1[\rulenameright{\oplus}]{\Gamma \modalsep \Theta \vdash \iota_i~V : A_1 \oplus A_2; \Delta}
\end{prooftree}} \\
\multicolumn{2}{c}{\begin{prooftree}
    \hypo{c_1 : (\Gamma, x : A_{\epsilon_1} \modalsep \Theta \vdash \Delta)}
    \hypo{c_2 : (\Gamma, y : B_{\epsilon_2} \modalsep \Theta \vdash \Delta)}
    \infer2[\rulenameleft{\oplus}]{\Gamma \modalsep \Theta ; \mumatch{x^{\epsilon_1}}{c_1}{y^{\epsilon_2}}{c_2} : A_{\epsilon_1} \oplus B_{\epsilon_2} \vdash \Delta}
\end{prooftree}} \\
\end{array} \vspace*{\fill} \mkern-12mu\begin{array}{cc}
\begin{prooftree}
    \hypo{\Gamma \modalsep \Theta \vdash V : A_{\positive}; \Delta}
    \infer1[\rulenameright{V}]{\Gamma \modalsep \Theta \vdash V : A_{\positive} \mid \Delta}
\end{prooftree} &
\begin{prooftree}
    \hypo{\Gamma \modalsep \Theta ; S : A_- \vdash \Delta}
    \infer1[\rulenameleft{S}]{\Gamma \modalsep \Theta \mid S : A_- \vdash \Delta}
\end{prooftree} \\
\multicolumn{2}{c}{\begin{prooftree}
    \hypo{c : (\Gamma \modalsep \Theta \vdash \alpha : A_{\epsilon_1}, \beta : B_{\epsilon_2}, \Delta)}
    \infer1[\rulenameright{\llpar}]{\Gamma \modalsep \Theta \vdash \mu(\alpha^{\epsilon_1}, \beta^{\epsilon_2}) . c : A_{\epsilon_2} \llpar B_{\epsilon_1}; \Delta}
\end{prooftree}} \\
\multicolumn{2}{c}{\begin{prooftree}
    \hypo{\Gamma \modalsep \Theta\mid e : A \vdash \Delta}
    \hypo{\Gamma' \modalsep \Theta' \mid f : B \vdash \Delta'}
    \infer2[\rulenameleft{\llpar}]{\Gamma, \Gamma' \modalsep \Theta, \Theta' \mid (e, f) : A \llpar B \vdash \Delta, \Delta'}
\end{prooftree}} \\
\mkern16mu\begin{prooftree}
    \hypo{c : (\Gamma, x : A_{\epsilon} \modalsep \Theta \vdash \Delta)}
    \infer1[\rulenameright{\neg}]{\Gamma \modalsep \Theta \vdash \mu[x^{\epsilon}] . c : \neg A_{\epsilon}; \Delta}
\end{prooftree}\mkern-16mu &
\begin{prooftree}
    \hypo{\Gamma \modalsep \Theta \vdash V : A; \Delta}
    \infer1[\rulenameleft{\neg}]{\Gamma \modalsep \Theta ; [V] : \neg A \vdash \Delta}
\end{prooftree} \\
\multicolumn{2}{c}{\begin{prooftree}
    \hypo{c_1 : (\Gamma, \modalsep \Theta \vdash \Delta, \alpha : A_{\epsilon_1}) }
    \hypo{c_2 : (\Gamma, \modalsep \Theta \vdash \Delta, \beta : B_{\epsilon_2}) }
    \infer2[\rulenameright{\with}]{\Gamma \modalsep \Theta \vdash \mustruct{\alpha^{\epsilon_1}}{c_1}{\beta^{\epsilon_2}}{c_2} : A_{\epsilon_1} \with B_{\epsilon_2}; \Delta}
\end{prooftree}} \\
\multicolumn{2}{c}{\begin{prooftree}
  \hypo{\Gamma \modalsep \Theta ; S : A_i^{\epsilon_i} \vdash \Delta}
  \infer1[\rulenameleft{\with}]{\Gamma \modalsep \Theta ; \pi_i~S : A_1^{\epsilon_1} \with A_2^{\epsilon_2} \vdash \Delta}
\end{prooftree}} \\
\begin{prooftree}
    \hypo{\Gamma_{\polbox} \modalsep \Theta \vdash V : A; \diamond}
    \infer1[\rulenameright{\typbox}]{\Gamma_{\polbox} \modalsep \Theta \vdash \valbox V : \typbox A; \diamond}
\end{prooftree} &
\begin{prooftree}
    \hypo{c : (\Gamma \modalsep x : A_{\epsilon}, \Theta \vdash \Delta)}
    \infer1[\rulenameleft{\typbox}]{\Gamma \modalsep \Theta ; \mutilde \valbox x^{\epsilon} . c &: \typbox A_{\epsilon} \vdash \Delta}
\end{prooftree} \\
\begin{prooftree}
    \hypo{\Gamma \modalsep \Theta \vdash V : A; \Delta}
    \infer1[\rulenameright{\mathfrak{R}(V)}]{\Gamma' \modalsep \Theta' \vdash \theta(V) : A ; \Delta'}
\end{prooftree} &
\begin{prooftree}
    \hypo{\Gamma \modalsep \Theta \vdash t : A \mid \Delta}
    \infer1[\rulenameright{\mathfrak{R}(t)}]{\Gamma' \modalsep \Theta' \vdash \theta(t) : A ; \Delta'}
\end{prooftree} \\
\begin{prooftree}
    \hypo{\Gamma \modalsep \Theta; S : A \vdash \Delta}
    \infer1[\rulenameleft{\mathfrak{R}(S)}]{\Gamma' \modalsep \Theta'; \theta(S) : A \vdash \Delta'}
\end{prooftree} &
\begin{prooftree}
    \hypo{\Gamma \modalsep \Theta \mid e : A \vdash \Delta}
    \infer1[\rulenameleft{\mathfrak{R}(e)}]{\Gamma' \modalsep \Theta' \mid \theta(e) : A \vdash \Delta'}
\end{prooftree} \\
\multicolumn{2}{c}{\begin{prooftree}
    \hypo{c : (\Gamma \modalsep \Theta \vdash \Delta)}
    \infer1[\rulename{\mathfrak{R}(c)}]{\theta(c) : (\Gamma' \modalsep \Theta' \vdash \Delta')}
\end{prooftree}}
\end{array}$
\end{small}
\caption{Typing rules of the system $\Lbox$}
\label{fig:lbox}
\end{figure*}

We note that some rule forces to separate the contexts, as the
R$\otimes$ rule. However, this is not an issue, because we can use
renamings to contract variables.
\begin{prooftree*}
  \infer0[\rulenameright{\text{Ax}}]{x : A \modalsep \diamond \vdash x : A; \diamond}
  \infer0[\rulenameright{\text{Ax}}]{y : A \modalsep \diamond \vdash y : A; \diamond}
  \infer2[\rulenameright{\otimes}]{x : A, y : A \modalsep \diamond \vdash (x, y) : A \otimes A; \diamond}
  \infer1[\rulenameright{\mathfrak{R}(V)}]{z : A \modalsep \diamond \vdash (z, z) : A \otimes A; \diamond}
\end{prooftree*}
We applied the renaming $\theta(x) = \theta(y) = z$. It preserves types : $x$ and $y$ are type $A$, and $z$ too.

Similarly, axiom rules only allow one variable/covariable in the context. Again, this is not an issue:
\begin{prooftree*}
  \infer0[\rulenameleft{\text{Ax}}]{\diamond \modalsep \diamond; \alpha : A \vdash \alpha : A}
  \infer1[\rulenameleft{\mathfrak{R}(S)}]{\Gamma \modalsep \Theta; \alpha : A \vdash \alpha : A, \Delta}
\end{prooftree*}
Here we applied the renaming $\theta(\alpha) = \alpha$. Of course, it preserves the type of $\alpha$. The other variables we added in the context do not appear in the expression, so nothing can go ``wrong'' with them.

Lastly, we can use renamings to exchange variables so that the context
are separated at the right place, for example, here, with $\theta(x) =
x$ and $\theta(y) = y$:
\begin{prooftree*}
  \infer0[\rulenameright{\typbox\text{Ax}}]{\diamond \modalsep y : B \vdash y : B; \diamond}
  \infer0[\rulenameright{\typbox\text{Ax}}]{\diamond \modalsep x : A \vdash x : A; \diamond}
  \infer2[\rulenameright{\otimes}]{\diamond \modalsep y : B, x : A \vdash (y, x) : B \otimes A; \diamond}
  \infer1[\rulenameright{\mathfrak{R}(V)}]{\diamond \modalsep x : A; y : B \vdash (y, x) : B \otimes A; \diamond}
\end{prooftree*}

This type system enjoys some soundness properties:

\begin{lemma}[Modal restriction]\label{th:modal-restriction}
  If $V$ is a value and $\Gamma \modalsep \Theta \vdash V :
  A_{\polbox} ; \Delta$ then $\Gamma_{\polbox} \modalsep \Theta \vdash
  V : A_{\polbox} ; \diamond$.
\end{lemma}

This is the crucial property of this calculus. It ensures that modal values only refer to modal values in their closure; thus, if we allocate modal values on the heap and the rest on the stack, it ensures that values on the heap do not refer to the stack.

\begin{lemma}[Typed substitution]
  If $c$ (or $t$, $V$, $S$ or $e$) is a well-typed command, expression, etc.
  in the context $(\Gamma \modalsep \Theta \vdash \Delta)$, if
  $(\Gamma' \modalsep \Theta' \vdash \Delta')$ is another context and
  $\sigma$ a function such that for all $x : A \in \Gamma$,
  $\sigma(x)$ is a value such that $\Gamma' \modalsep \Theta' \vdash
  \sigma(x) : A; \Delta'$, for each $x : A \in \Theta$, $\sigma(x)$ is
  a value such that $\Gamma'_{\polbox} \modalsep \Theta' \vdash
  \sigma(x) : A; \diamond$ and for each $\alpha : A \in \Delta$,
  $\sigma(\alpha)$ is a co-value such that $\Gamma' \modalsep \Theta';
  \sigma(\alpha) : A \vdash \Delta'$, then $\sigma(c)$, respectively
  $\sigma(t)$, etc. is well-typed in the context $(\Gamma' \modalsep
  \Theta' \vdash \Delta')$. Moreover, if applicable, its type is the
  same.
\end{lemma}

\begin{lemma}[Subject reduction]\label{th:subject-reduction}
  If $c \red c'$ and $c : (\Gamma \modalsep \Theta \vdash \Delta)$
  then $c' : (\Gamma \modalsep \Theta \vdash \Delta)$.
\end{lemma}

\begin{lemma}[Normalization]\label{th:strong-normalization}
  $\red$ is normalizing on well-typed terms.
\end{lemma}

\begin{proof}
  We have in fact the stronger result: the contextual closure of
  $\red$ is strongly normalizing. Indeed, $\Lbox$ can be translated
  into the $\mathrm{LJ}^{\eta}_p$ system from
  \citet{curien_theory_2016} in a straightforward manner, by
  translating $\typbox A$ as $\Upshift A$, $(\Gamma \modalsep \Theta
  \vdash \Delta)$ as $(\Gamma, \Upshift \Theta \vdash \Delta)$,
  $\typbox V$ as $(V, ())$ and $\mutilde \valbox x . c$ as $\mutilde
  \Upshift x . c$. Moreover, the polarity $\polbox$ is translated as
  $+$. This translation preserves typing and is a simulation, and
  their system is strongly normalizing (see e.g.
  \citep{munch-maccagnoni_note_2017}); hence ours is also strongly
  normalizing.
\end{proof}

Together, these lemmas allow to prove that we can evaluate every
``closed'' command $c : (\diamond \modalsep \diamond \vdash \tp :
R_{\polbox})$ to a value, where $R$ is a modal type and $\tp$ the
toplevel continuation. Indeed, such a command reduces to a normal form
$c'$. By case analysis, we can see that $c' =
\command{\polbox}{V}{\tp}$ for a certain $V$. This $V$ has type $R$ in
the context $(\diamond \modalsep \diamond \vdash \tp : R)$ by subject
reduction, but thanks the modal restriction lemma, we can actually
show that it contains no free variable. This $V$ is unique because
$\red$ is deterministic.

\begin{theorem}[Evaluation]\label{th:eval}
  If $c : (\diamond \modalsep \diamond \vdash \tp : R_{\polbox})$
  there exists a closed value $V$ of type $R$ such that $c \red^*
  \command{\polbox}{V}{\tp}$. Moreover, this $V$ is unique.
\end{theorem}

This is interesting because in the end, we are interested in
evaluating complete programs that return purely positive values like
booleans or integers; which are indeed given by modal datatypes (e.g.
$\unit \oplus \unit$ which is a modal type). Like the type system of
$\lambda_{1/2}$ described in \cite{cong_compiling_2019}, in this
system, we can wrap a type in $\typbox$ to force its values to be
pure, ie., not to contain continuations, but we may locally use
continuations in a function which returns a pure value, and the type
system ensures that continuations do not escape.

\section{Machine-like semantics} \label{sec:machine}

This section presents a machine-like semantics for the calculus which
manages the memory explicitly. For their calculus, \citet{cong_compiling_2019} devised a machine with an interesting property, stackability: when calling a function that accepts a first-class argument, the stack gets freed and reset to how it was at the definition of the function. Here, we do not have the notion of
first-class or second-class types directly, but a first-class type in
their system corresponds to a modal type in ours, so a function
accepting a first-class argument would have type $A_{\polbox} \to
B_{\polbox} = \Upshift (\neg A_{\polbox} \llpar B_{\polbox})$
in ours.
Note that $B$ is always first-class in their setting, so modal in
ours. The semantics we propose is based on generalizing the
stackability property.

We first present the memory of our machine in
\cref{gra:machine-syntax}. It can store values and co-values. The
typing rules will ensure that whenever a variable $x$ is stored in
$\envM$, then it must be negative, ie. represent a thunk which has not
yet been evaluated. More generally, if a positive or modal variable
$x$ appears in a value $V$ stored in the memory, then every occurence
of $x$ appears as a subterm of some command $c$, itself a subterm of
$V$. Dually, negative covariables appearing in a co-value $S$ stored
inside $\envM$ can only appear as subterms of some command which is a
subterm of $S$. Thus, if $x$ is positive, then the strict pair $(x,
x)$ can not be stored in $\envM$, but
$\mustruct{\alpha^+}{\command{+}{x}{\alpha}}{\beta^+}{\command{+}{x}{\beta}}$,
representing the lazy record both of whose fields have value $x$, can
be stored in $\envM$. Dually, stored covariables $\alpha$ must be
positive or modal. To outline this distinction, we will write $\semV,
\semW$ or $\semS$ for values and co-values stored inside $\envM$.

\begin{grammar}[Memory syntax][][gra:machine-syntax]
  \firstcasesubtil{$\envH$}{\diamond \gralt \envH, x^{\epsilon} \coloneq \semV}{Heap}
  \firstcasesubtil{$\envE$}{x^{\notmodal} \coloneq \semV \gralt \alpha^{\epsilon} \coloneq \semS}{Equations}
  \firstcasesubtil{$\envF$}{(\envE) \gralt (\envE, \envE)}{Stack frame}
  \firstcasesubtil{$\envS$}{\diamond \gralt \envS, \envF}{Stack}
  \firstcasesubtil{$\envM$}{\mem{\envH}{\envS}}{Memory}
\end{grammar}

$\envM$ is the memory; it is composed of a heap, $\envH$, for values
of modal types, and of a stack $\envS$, for values of non-modal types and
for co-values. We can also allocate values of non-modal type in $\envH$, but only if they don't refer to variables outside of $\Theta$ except for variables of modal type. This ensures that values on the heap don't refer to the stack. The stack is divided into stack frames $\envF$.
We can assign new values with the operations
$\envM[x^{\epsilon} \coloneq \semV]$ and $\envM[\alpha^{\epsilon} \coloneq \semS]$,
which allocates the memory on the heap or the
stack accordingly: each time new variables are allocated on the stack, a new stack frame is created. A stack frame has an address: we denote $@x$ or $@\alpha$ the adress of the stack frame in which $x$ or $\alpha$ has been allocated. We can restrict the memory $\restrict{\envM}{@\alpha}$, so that it drops the stack frame to which $\alpha$ belongs, and every stack frame allocated afterwards. We write $\envM(x)$ and
$\envM(\alpha)$ for the value, respectively the stack, assigned to $x$
and $\alpha$ inside $\envM$, and $\varpi(x)$ or $\varpi(\alpha)$ for
its polarity $\epsilon$ (values on the heap are all modal). The memory
will be implicit in the notation.

More formally, for the allocation:
\[\begin{array}{ccc}
  \mem{\envH}{\envS}[x^{\polbox} \coloneq \semV] & \defeq & \mem{\envH, (x^{\polbox} \coloneq \semV)}{\envS} \\
  \mem{\envH}{\envS}[x^{\notmodal} \coloneq \semV] & \defeq & \mem{\envH}{\envS, (x^{\epsilon} \coloneq \semV)} \\
  \mem{\envH}{\envS}[\alpha^{\epsilon} \coloneq \semS] & \defeq & \mem{\envH}{\envS, (\alpha^{\epsilon} \coloneq \semS)} \\
  \mem{\envH}{\envS}[x^{\polbox} \coloneq \semV, y^{\polbox} \coloneq \semV'] & \defeq & \mem{\envH, (x \coloneq \semV), (y \coloneq \semV')}{\envS} \\
  \mem{\envH}{\envS}[x^{\polbox} \coloneq \semV, \alpha^{\epsilon} \coloneq \semS] & \defeq & \mem{\envH, (x \coloneq \semV)}{\envS, (\alpha^{\epsilon} \coloneq \semS)} \\
  \mem{\envH}{\envS}[\alpha^{\epsilon} \coloneq \semS, \beta^{\epsilon'} \coloneq \semS'] & \defeq & \mem{\envH}{\envS, (\alpha^{\epsilon} \coloneq \semS, \beta^{\epsilon'} \coloneq \semS')} \\
  & \text{etc.} &
\end{array}\]
And for the restriction:
\[\begin{array}{ccc}
  \restrict{\mem{\envH}{\diamond}}{@\alpha} & \defeq & \mem{\envH}{\diamond} \\
  \restrict{\mem{\envH}{\envS, (\alpha^{\epsilon} \coloneq \semS)}}{@\alpha} & \defeq & \mem{\envH}{\envS} \\
  \restrict{\mem{\envH}{\envS, (\beta^{\epsilon} \coloneq \semS)}}{@\alpha} & \defeq & \restrict{\mem{\envH}{\envS}}{@\alpha} \\
  \restrict{\mem{\envH}{\envS, (x^{\notmodal} \coloneq \semV)}}{@\alpha} & \defeq & \restrict{\mem{\envH}{\envS}}{@\alpha} \\
  \restrict{\mem{\envH}{\envS, (\beta^{\epsilon} \coloneq \semS, \gamma^{\epsilon'} \coloneq \semS')}}{@\alpha} & \defeq & \restrict{\mem{\envH}{\envS}}{@\alpha} \\
  \restrict{\mem{\envH}{\envS, (\alpha^{\epsilon} \coloneq \semS, \beta^{\epsilon'} \coloneq \semS')}}{@\alpha} & \defeq & \mem{\envH}{\envS} \\
  \restrict{\mem{\envH}{\envS, (\beta^{\epsilon'} \coloneq \semS, \alpha^{\epsilon} \coloneq \semS')}}{@\alpha} & \defeq & \mem{\envH}{\envS} \\
  & \text{etc.} &
\end{array}\]
Now, we show how to evaluate values and co-values. We define the following
relations (\cref{fig:semvalstack}):
\[ \evalval{\envM}{V}{\semV} \qquad \evalstack{\envM}{S}{\semS} \]
The first one means that $V$ evaluates to the value $\semV$ in the
environment $\envM$, and the second means that $S$ evaluates to the
co-value $\semS$ in the environment $\envM$. They do not force thunks;
thus, negative variables are not evaluated, and similarly for positive
or modal covariables.

\begin{figure}[h]
\renewcommand{\arraystretch}{2.7}
\renewcommand{\arraycolsep}{1.5ex}
\(\begin{array}{cc}
\begin{prooftree}
  \hypo{\envM(x) = \semV}
  \hypo{\varpi(x) \in \{+, \polbox\}}
  \infer2{\evalval{\envM}{x}{\semV}}
\end{prooftree}
&
\begin{prooftree}
  \hypo{\varpi(x) = -}
  \infer1{\evalval{\envM}{x}{x}}
\end{prooftree} \\

\begin{prooftree}
  \hypo{\varpi(\alpha) \in \{+, \polbox\}}
  \infer1{\evalstack{\envM}{\alpha}{\alpha}}
\end{prooftree}
&
\begin{prooftree}
  \hypo{\envM(\alpha) = \semS}
  \hypo{\varpi(\alpha) = -}
  \infer2{\evalstack{\envM}{\alpha}{\semS}}
\end{prooftree} \\

\begin{prooftree}
  \hypo{\evalval{\envM}{V}{\semV}}
  \hypo{\evalval{\envM}{V'}{\semW}}
  \infer2{\evalval{\envM}{(V, V')}{(\semV, \semW)}}
\end{prooftree}
&
\begin{prooftree}
  \hypo{\evalval{\envM}{V}{\semV}}
  \infer1{\evalval{\envM}{\iota_i~V}{\iota_i~\semV}}
\end{prooftree} \\

\begin{prooftree}
  \infer0{\evalstack{\envM}{\mutilde(x^{\epsilon_1}, y^{\epsilon_2}) . c}{\mutilde(x^{\epsilon_1}, y^{\epsilon_2}) . c}}
\end{prooftree}
&
\begin{prooftree}
  \hypo{\evalstack{\envM}{S}{\semS}}
  \infer1{\evalstack{\envM}{\pi_i~S}{\pi_i~\semS}}
\end{prooftree} \vspace*{-1ex}\\

\multicolumn{2}{c}{\begin{prooftree}
  \infer0{\evalstack{\envM}{\mumatch{x_1^{\epsilon_1}}{c_1}{x_2^{\epsilon_2}}{c_2}}{\mumatch{x_1^{\epsilon_1}}{c_1}{x_2^{\epsilon_2}}{c_2}}}
\end{prooftree}} \vspace*{-1ex}\\

\multicolumn{2}{c}{\begin{prooftree}
  \infer0{\evalval{\envM}{\mustruct{\alpha_1^{\epsilon_1}}{c_1}{\alpha_2^{\epsilon_2}}{c_2}}{\mustruct{\alpha_1^{\epsilon_1}}{c_1}{\alpha_2^{\epsilon_2}}{c_2}}}
\end{prooftree}} \vspace*{-1ex}\\

\begin{prooftree}
  \infer0{\evalval{\envM}{()}{()}}
\end{prooftree}
&
\begin{prooftree}
  \infer0{\evalstack{\envM}{\mutilde () . c}{\mutilde () . c}}
\end{prooftree} \\

\begin{prooftree}
  \hypo{\evalval{\envM}{V}{\semV}}
  \infer1{\evalstack{\envM}{[V]}{[\semV]}}
\end{prooftree}
&
\begin{prooftree}
  \infer0{\evalval{\envM}{\mu[x^{\epsilon}] . c}{\mu[x^{\epsilon}] . c}}
\end{prooftree} \\

\begin{prooftree}
  \hypo{\evalstack{\envM}{S}{\semS}}
  \hypo{\evalstack{\envM}{S'}{\semS'}}
  \infer2{\evalstack{\envM}{(S, S')}{(\semS, \semS')}}
\end{prooftree}
&
\begin{prooftree}
  \infer0{\evalval{\envM}{\mu(\alpha^{\epsilon_1}, \beta^{\epsilon_2}) . c}{\mu(\alpha^{\epsilon_1}, \beta^{\epsilon_2}) . c}}
\end{prooftree} \\

\begin{prooftree}
  \hypo{\evalval{\envM}{V}{\semV}}
  \infer1{\evalval{\envM}{\valbox V}{\valbox \semV}}
\end{prooftree}
&
\begin{prooftree}
  \infer0{\evalstack{\envM}{\mutilde\valbox x^{\epsilon} . c}{\mutilde\valbox x^{\epsilon} . c}}
\end{prooftree}
\end{array}\)
\caption{Evaluation of values and stacks}
\label{fig:semvalstack}
\end{figure}

For example, we have:
\[ \evalval{\mem{x \coloneq ()}{(y^- \coloneq \semV)}}{(x, y)}{((), y)} \]

Now, we define in \cref{fig:semcmd} the relation
$\redcmd{\envM}{c}{c'}{\envM'}$, which means that with the memory
$\envM$, the command $c$ reduces to the command $c'$ and modifies the
memory to be $\envM'$. The key point of this semantics are the rule Eval$\mutilde\valbox$, which allocates modal variables (belonging to $\Theta$) on the heap:
\begin{prooftree*}
  \hypo{\evalval{\envM}{V}{\typbox \semV}}
  \infer1[Eval$\mutilde\valbox$]{\redcmd{\mem{\envH}{\envS}}{\command{\polbox}{V}{\mutilde \valbox x^{\epsilon} . c}}{c}{\mem{\envH, (x^{\epsilon} \coloneq \semV)}{\envS}}}
\end{prooftree*}
and the rule Eval$\typbox$:
\begin{prooftree*}
  \hypo{\envM(\alpha) =^{\polbox} \semS}
  \infer1[Eval$\typbox$]{\redcmd{\envM}{\command{\polbox}{V}{\alpha}}{\command{\polbox}{V}{\semS}}{\restrict{\envM}{@\alpha}}}
\end{prooftree*}
After evaluating a covariable $\alpha$ of modal type, we can
make the stack shrink. Indeed, if we restrict ourselves to
typed values, $V$ is of modal type, so it cannot contains
references to $\envS$; and the value $\semS$ of $\alpha$ must
be typed in the environment we had before $\alpha$ was allocated.

\begin{figure*}
\begin{small}
\renewcommand{\arraystretch}{2.9}
\arraycolsep=0ex
\(\begin{array}{cc}
\begin{prooftree}
  \hypo{\envM(\alpha) =^+ \semS}
  \infer1[Eval$+$]{\redcmd{\envM}{\command{+}{V}{\alpha}}{\command{{+}}{V}{\semS}}{\envM}}
\end{prooftree}
&
\begin{prooftree}
  \hypo{\envM(x) =^- \semV}
  \infer1[Eval$-$]{\redcmd{\envM}{\command{-}{x}{S}}{\command{-}{\semV}{S}}{\envM}}
\end{prooftree} \\

\begin{prooftree}
  \hypo{\envM(\alpha) =^{\polbox} \semS}
  \infer1[Eval$\typbox$]{\redcmd{\envM}{\command{\polbox}{V}{\alpha}}{\command{\polbox}{V}{\semS}}{\restrict{\envM}{@\alpha}}}
\end{prooftree} &

\begin{prooftree}
  \hypo{\evalstack{\envM}{S}{[\semV]}}
  \infer1[Eval$\mu\neg$]{\redcmd{\envM}{\command{-}{\mu[x^\epsilon] . c}{S}}{c}{\envM[x^\epsilon \coloneq \semV]}}
\end{prooftree} \\

\begin{prooftree}
  \hypo{\evalstack{\envM}{S}{\semS}}
  \infer1[Eval$\mu\alpha^{\epsilon}$]{\redcmd{\envM}{\command{\epsilon}{\mu \alpha^\epsilon . c}{S}}{c}{\envM[\alpha^\epsilon \coloneq \semS]}}
\end{prooftree} &

\begin{prooftree}
  \hypo{\evalval{\envM}{V}{\semV}}
  \infer1[Eval$\mutilde x^{\epsilon}$]{\redcmd{\envM}{\command{\epsilon}{V}{\mutilde x^\epsilon . c}}{c}{\envM[x^\epsilon \coloneq \semV]}}
\end{prooftree} \\

\multicolumn{2}{c}{\begin{prooftree}
  \hypo{\evalval{\envM}{V}{(\semV, \semV')}}
  \infer1[Eval$\mutilde\otimes$]{\redcmd{\envM}{\command{\epsilon}{V}{\mutilde(x^{\epsilon_1}, y^{\epsilon_2}) . c}}{c}{\envM[x^{\epsilon_1} \coloneq \semV, y^{\epsilon_2} \coloneq \semV']}}
\end{prooftree}} \\

\multicolumn{2}{c}{\begin{prooftree}
  \hypo{\evalstack{\envM}{S}{(\semS, \semS')}}
  \infer1[Eval$\mu\llpar$]{\redcmd{\envM}{\command{-}{\mu(\alpha^{\epsilon_1}, \beta^{\epsilon_2}) . c}{S}}{c}{\envM[\alpha^{\epsilon_1} \coloneq \semS, \beta^{\epsilon_2} \coloneq \semS']}}
\end{prooftree}} \\

\multicolumn{2}{c}{\begin{prooftree}
  \hypo{\evalval{\envM}{V}{\iota_i~\semV}}
  \infer1[Eval$\mutilde\oplus$]{\redcmd{\envM}{\command{\epsilon}{V}{\mumatch{x_1^{\epsilon_1}}{c_1}{x_2^{\epsilon_2}}{c_2}}}{c_i}{\envM[x_i^{\epsilon_i} \coloneq \semV]}}
\end{prooftree}} \\

\multicolumn{2}{c}{\begin{prooftree}
  \hypo{\evalstack{\envM}{S}{\pi_i~\semS}}
  \infer1[Eval$\mu\with$]{\redcmd{\envM}{\command{-}{\mustruct{\alpha_1^{\epsilon_1}}{c_1}{\alpha_2^{\epsilon_2}}{c_2}}{S}}{c_i}{\envM[\alpha_i^{\epsilon_i} \coloneq \semS]}}
\end{prooftree}} \\

\begin{prooftree}
  \hypo{\evalval{\envM}{V}{()}}
  \infer1[Eval$\mutilde\unit$]{\redcmd{\envM}{\command{\polbox}{V}{\mutilde () . c}}{c}{\envM}}
\end{prooftree} &

\begin{prooftree}
  \hypo{\evalval{\envM}{V}{\typbox \semV}}
  \infer1[Eval$\mutilde\valbox$]{\redcmd{\mem{\envH}{\envS}}{\command{\polbox}{V}{\mutilde \polbox x^{\epsilon} . c}}{c}{\mem{\envH, x^{\epsilon} \coloneq \semV}{\envS}}}
\end{prooftree} \\

\end{array}\)
\end{small}

\caption{Evaluation of commands}
\label{fig:semcmd}
\end{figure*}

Let us consider what happens when evaluating, for example, a (CBV)
function $t$ on some expression $u$. We suppose that $t$ has for type
$\Upshift (A_{\polbox} \to B_{\polbox})$, that is, where $A$ and $B$
are modal. Note that $t$ is not of modal type. This is the situation
considered in \citet{cong_compiling_2019}. Now consider the evaluation
of the following expression:
\[ \mu\gamma^{\polbox} . \command{+}{t}{\mutilde f^+ . \command{\polbox}{\mu\alpha^{\polbox} . \command{\polbox}{u}{\alpha}}{\mutilde x^{\polbox} . \command{+}{f^+}{\mutilde \Upshift g^- . \command{-}{g}{x \cdot \gamma}}}} \]
First, we must evaluate $t$. Suppose it evaluates to $f$. Then we
evaluate $u$ (note the $\eta$-expansion). Suppose it returns a value
$V$: this value will be evaluated against $\alpha$, appearing in
$\command{\polbox}{V}{\alpha}$. Thus, at this point, because $\alpha$
is modal, the stack will be reset a first time to the point where it
was before we started evaluating $u$. This is valid because $V$
contains no reference to the stack, as it is modal, and $\alpha$ points
to $\mutilde x^{\polbox} . \command{+}{f^+}{\mutilde \Upshift g^- .
  \command{-}{g}{x \cdot \gamma}}$, whose free variables have been
allocated before the beginning of the evaluation of $u$. Then, we store $x
\coloneq V$ on the heap, we extract the real function $g$ from its
wrapper $f$ and evaluate it on $x$. This returns some value $V'$,
which is evaluated against $\gamma$. At this point, because $\gamma$
is also modal, we can restore the stack to how it was before the call
of $t~u$: thus, the stack is restored in its initial state, and no
memory has been leaked. Even if $A$ is non-modal, we can still ensure
that the stack is freed when the function returns, though we cannot
free it after evaluating $u$. This is exactly the stack discipline of
the CPS translation presented in \citet{cong_compiling_2019}.

More generally, in this semantics, co-values $S$ and non-modal values $V$ follow a stack discipline:

\begin{theorem}[Stackability]
The semantics that allocates co-values $S$ and non-modal values $V$ on the stack $\envS$, shrinking every time the evaluation of a modal expression returns, bringing the stack back in its state before the evaluation of the modal expression started, is sound.
\end{theorem}

In particular, this applies to functions whose return type is modal,
and to their arguments if they are modal. That means that the stack in
which we allocate correspond to the call stack if we restrict the
calculus to functions whose return type is modal.

However, the behaviour of functions returning non-modal types is a bit weird: for example, $(\lambda x^+ . \letin{y}{(x, x)}{y})\;V$ is translated as $t\defeq\mu\beta^+ . \command{-}{\mu(x \cdot \alpha) . \command{+}{(x, x)}{\mutilde y^+ . \command{+}{y}{\alpha}}}{V \cdot \beta}$, for which we have:
\begin{multline*}
  \left(\mem{\envH}{\envS} \vDash \command{+}{t}{S}\right) \rightsquigarrow^*
  \left(\mem{\envH}{\envS, (\beta^+ \coloneqq \semS), (x^+ \coloneq \semV, \alpha^+ \coloneqq \beta), (y^+ \coloneq (x, x))} \vDash \command{+}{y}{\alpha}\right)
\end{multline*}
where $S$ is an arbitrary co-value of positive type, and where we have
$\evalstack{\mem{\envH}{\envS}}{S}{\semS}$ and
$\evalval{\mem{\envH}{\envS, (\beta^+ \coloneqq \semS)}}{V}{\semV}$.
That is, the returned value, $y$ depends on the last stack frame,
especially on the binding $y^+ \coloneqq (x, x)$, which was not
allocated at the time the evaluation started. Thus the stack cannot
shrink and return to $\envS$ at this point of the execution.

In particular, this means that without the full second-class
restriction, when functions can return non-modal values, the stack in
which we allocate does not correspond to a call stack. On the one
hand, whether and how the stackability property allows us to improve
over indiscriminate heap-allocation remains to be investigated. On the
other hand, this is not necessary to our point that the metatheory of
second-class continuations follows from that of S4 sequent
calculus---and the latter might admit more relaxed characterisations
of the notion of non-escaping values (in the same way as second-class
values are a simplistic form of borrowing).

\section{Correctness}

We now prove the correctness of this machine-like semantics with
respect to the operational semantics $\red$.

We first define a typing relation for environments in
\cref{fig:typmem}. We first fix a modal type $R$, which will not
change along this work, and will serve as a return type, and the type
of $\tp$. Type environments are defined almost like environments,
except that we replace $x^\epsilon \coloneq \semV$ by $x :
A_{\epsilon}$, etc. We note such a type environment $\Sigma$ or
$\mem{\Sigma_{\envH}}{\Sigma_{\envS}}$. We can associate a typing
context to a type environment: variables in $\envS$ go into $\Gamma$,
covariables go into $\Delta$ and variables in $\envH$ go into
$\Theta$. We also always add $\tp : R$ to $\Delta$. For example,
\[
\sem{\mem{(x : A), (y : B)}{(\alpha : C, \beta : D), (z : E)}} = z :
E \modalsep x : A, y : B \vdash \tp : R, \alpha : C, \beta : B
\]
Note
that our rules ensure that a value or a stack allocated on the stack
only contains references to the heap and to variables allocated
before. \Cref{th:modal-restriction} ensures that values on the heap
only refer to the heap.

\begin{figure}
\begin{small}
\renewcommand{\arraystretch}{3.5}
\renewcommand{\arraycolsep}{0em}
\vspace*{-3ex}
\(\begin{array}{c}
\begin{prooftree}
  \infer0{\mem{\diamond}{\diamond} : \mem{\diamond}{\diamond}}
\end{prooftree} \vspace*{-2ex}\\

\begin{prooftree}
  \hypo{\mem{\envH}{\envS} : \mem{\Sigma_{\envH}}{\Sigma_{\envS}}}
  \hypo{\Gamma_{\polbox} \modalsep \Theta \vdash \semV : A_{\epsilon}; \diamond}
  \hypo{\sem{\mem{\envH}{\envS}} = (\Gamma \modalsep \Theta \vdash \Delta)}
  \infer3{\mem{\envH, (x^{\epsilon} \coloneq \semV)}{\envS} : \mem{\Sigma_{\envH}, (x : A_{\epsilon})}{\Sigma_{\envS}}}
\end{prooftree} \\

\begin{prooftree}
  \hypo{\mem{\envH}{\envS} : \mem{\Sigma_{\envH}}{\Sigma_{\envS}}}
  \hypo{\Gamma \modalsep \Theta \vdash \semV : A_{\notmodal}; \Delta}
  \hypo{\sem{\mem{\envH}{\envS}} = (\Gamma \modalsep \Theta \vdash \Delta)}
  \infer3{\mem{\envH}{\envS, (x^{\notmodal} \coloneq \semV)} : \mem{\Sigma_{\envH}}{\Sigma_{\envS}, (x : A_{\notmodal})}}
\end{prooftree} \\

\begin{prooftree}
  \renewcommand{\arraystretch}{1}
  \hypo{\mem{\envH}{\envS} : \mem{\Sigma_{\envH}}{\Sigma_{\envS}}}
  \hypo{\begin{array}{c}
    \Gamma \modalsep \Theta; \semS : A_{\epsilon} \vdash \Delta \\
    \Gamma \modalsep \Theta; \semS' : B_{\epsilon'} \vdash \Delta
  \end{array}}
  \hypo{\sem{\mem{\envH}{\envS}} = (\Gamma \modalsep \Theta \vdash \Delta)}
  \infer3{\mem{\envH}{\envS, (\alpha^{\epsilon} \coloneq \semS, \beta^{\epsilon'} \coloneq \semS')} : \mem{\Sigma_{\envH}}{\Sigma_{\envS}, (\alpha : A_{\epsilon}, \beta : B_{\epsilon'})}}
\end{prooftree}
\end{array}\)

\vspace*{1em}\text{etc.}
\end{small}

\caption{Typing of environments}
\label{fig:typmem}
\end{figure}

We enumerate some useful properties of this type system:

\begin{lemma}[Progress]\label{th:machine-progress}
  If $\envM : \Sigma$ and $c$ is well-typed in the environment
  $\sem{\Sigma}$ then $c = \command{+}{\semV}{\tp}$ or there exists
  $c'$ and $\envM'$ such that $\redcmd{\envM}{c}{c'}{\envM'}$.
\end{lemma}

\begin{lemma}[Subject reduction]\label{th:machine-subject}
  If $\envM : \Sigma$ and $c$ is well-typed in the environment
  $\sem{\Sigma}$ and $\redcmd{\envM}{c}{c'}{\envM'}$, then $\envM'$ is
  typable, either in an extension or in a restriction $\Sigma'$ of
  $\Sigma$, and $c'$ is well-typed in the environment $\Sigma'$.
\end{lemma}

\begin{proof}[Proof sketch]
  By case analysis on the derivation $\redcmd{\envM}{c}{c'}{\envM'}$.
  For the case of the rule $Eval\typbox$, we use the
  \cref{th:modal-restriction} to prove that $V$ contains no reference
  to $\envS$; and then, if $\envM(\alpha) =^{\polbox} \semS$ and $\envM$
  is well-typed, then $\semS$ must be typed in
  $\sem{\restrict{\envM}{@\alpha}}$. Thus
  $\command{\polbox}{V}{\semS}$ is well-typed in
  $\sem{\restrict{\envM}{@\alpha}}$.
\end{proof}

\begin{lemma}
  If $\envM : \Sigma$ then $\envM$ defines a typed substitution from
  $\sem{\Sigma}$ to $\sem{\mem{\diamond}{\diamond}}$, which will be
  noted $[\envM]$.
\end{lemma}

\begin{lemma}
  If $\envM : \Sigma$ and $V$ has type $A$ in the environment
  $\sem{\Sigma}$ and $\evalval{\envM}{V}{\semV}$ then $V[\envM] =
  \semV[\envM]$ ; the same goes for $\semS$.
\end{lemma}

\begin{lemma}\label{th:machine-simulation}
  If $\envM : \Sigma$ and $c$ is well-typed in the environment
  $\sem{\Sigma}$ and $\redcmd{\envM}{c}{c'}{\envM'}$ then $c[\envM]
  \red^{\leq 1} c'[\envM']$. Moreover, there cannot be an infinite
  sequence of $\rightsquigarrow$ that yields an equality.
\end{lemma}

Now, if $c$ is closed, ie. defined in the environment
$\sem{\mem{\diamond}{\diamond}} = (\diamond \modalsep \diamond \vdash
\tp : R)$, by \cref{th:eval} there exists a unique $V$ of type $R$
such that $c \red^* \command{+}{V}{\tp}$. We will show the
following:

\begin{theorem}[Evaluation]
  There exists $V'$ and $\envH$ such that
  $\redcmd*{\mem{\diamond}{\diamond}}{c}{\command{\polbox}{V'}{\tp}}{\mem{\envH}{\envS}}$, $V'$ contain no reference to $\envS$ and $V'[\envH] = V$.
\end{theorem}

This proves the soundness of our semantics.

\begin{proof}[Proof sketch]
  With the lemmas proved before, we can show that $\rightsquigarrow$
  is normalizing. Moreover, at the end, the resulting
  command must be of the form $\command{\polbox}{V'}{\tp}$, in some
  environment $\envM'$. But $V'$ is a value of modal type, so it
  cannot contain references to any variables or covariables in
  $\envS'$ or to $\tp$, by \cref{th:modal-restriction}.
\end{proof}

\printbibliography

\end{document}